\documentclass[aps,preprint,prb,superscriptaddress]{revtex4}
\usepackage{amsfonts}
\usepackage[final,hiresbb]{graphicx}
\usepackage{amsmath}
\usepackage{amssymb}
\usepackage{amstext}

\usepackage{color}
\usepackage{braket}

\usepackage{subfigure}

\definecolor{Green}{RGB}{0,204,102}
\definecolor{Purple}{RGB}{102,0,255}
\definecolor{Blue}{RGB}{0,0,255}
\definecolor{Red}{RGB}{151,010,010}

\begin{document}

\title{Electric Dipole Coupling in Optical Cavities and Its Implications for Energy Transfer, Upconversion and Pooling} 

\author{Michael D. LaCount}
\affiliation{Department of Physics, Colorado School of Mines, Golden, CO 80401, USA}
\author{Mark T. Lusk}
\email{mlusk@mines.edu}
\affiliation{Department of Physics, Colorado School of Mines, Golden, CO 80401, USA}

\keywords{quantum interference, exciton, transistor, quantum interference, pulse shaping, transport, mesoscopic, quantum control}

\begin{abstract}
Resonant energy transfer, energy transfer upconversion, and energy pooling are considered within optical cavities to elucidate the relationship between exciton dynamics and donor/acceptor separation distance. This is accomplished using perturbation theory to derive analytic expressions for the electric dipole coupling tensors of perfect planar and rectangular channel reflectors---directly related to a number of important energy transfer processes. In the near field, the separation dependence along the cavity axis is not influenced by the cavity and is essentially the same as for three-dimensional, free space. This is in sharp contrast to the reduced sensitivity to separation found in idealized low-dimensional settings. The cavity dynamics only correspond to their reduced dimensional counterparts in the far field where such excitonic processes are not typically of interest. There is an intermediate regime, though, where sufficiently small cavities cause a substantial decrease in separation sensitivity that results in one component of the dipole-dipole coupling tensor being much larger than those of free space.
\end{abstract}

\maketitle

\newpage
{\bf Keywords}: cavity, resonant energy transfer, upconversion, organic, exciton, photon, quantum electrodynamics, superradiance

\section{INTRODUCTION}

The energy of light absorbed by a semiconducting nanostructure manifests itself as an electron-hole pair, an exciton, that can subsequently engage in a rich variety of relaxation processes. It may couple to phonon modes and simply generate heat or re-emerge as a photon through photoluminescence. In the vicinity of another nanostructure, though, the exciton may also move from one site to another~\cite{Forster} without any exchange of electrons~\cite{Dexter}. Such a process emerges from a superposition of virtual photon interactions between donor and acceptor moieties and is commonly referred to as Resonance Energy Transfer (RET). It underlies an impressive array of natural and technological processes and has been studied extensively for many decades. 

Beyond simple RET, interactions between multiple excitons can result in the creation of a single, high-energy excitation. Such upconversions are referred to as \emph{Energy Transfer Upconversion} (ETU)\cite{Wang09} if the intensified exciton is created on one of the original sites and \emph{Energy Pooling} (EP) if it emerges on a third nanostructure~\cite{Jenkins, Lacount_2015, Liang_2009, Pushkara_2011}. These processes, for instance, allow the infrared spectrum to be drawn upon to carry out higher energy tasks in photovoltaics\citep{Xie_2012,Zou_2012,Trupke_2006}, biofuel production~\cite{Wondraczek_2013} and medical applications\cite{Ang_2011,Chatterjee_2010, Theranostics_Chen_2014, Cancer_Dai_2013}.

Motivated by a desire to make RET and ETU more robust and efficient, optical cavities have been experimentally realized in which these processes occur. Cavities offer a higher density of optical states at resonance which enhances the efficiency of RET relative to photoluminescence~\cite{Hopmeier_1999}. For the same reasons, the efficiency of ETU was was found to increase by two orders of magnitude in cavity environment\cite{Xu_2014}. Early indications that cavities also increased the rate of RET were ultimately discounted though~\cite{Blum_2012}. 

\begin{figure}[h]
\centering{}
\subfigure[ ]{\centering{}\includegraphics[width=0.40\textwidth]{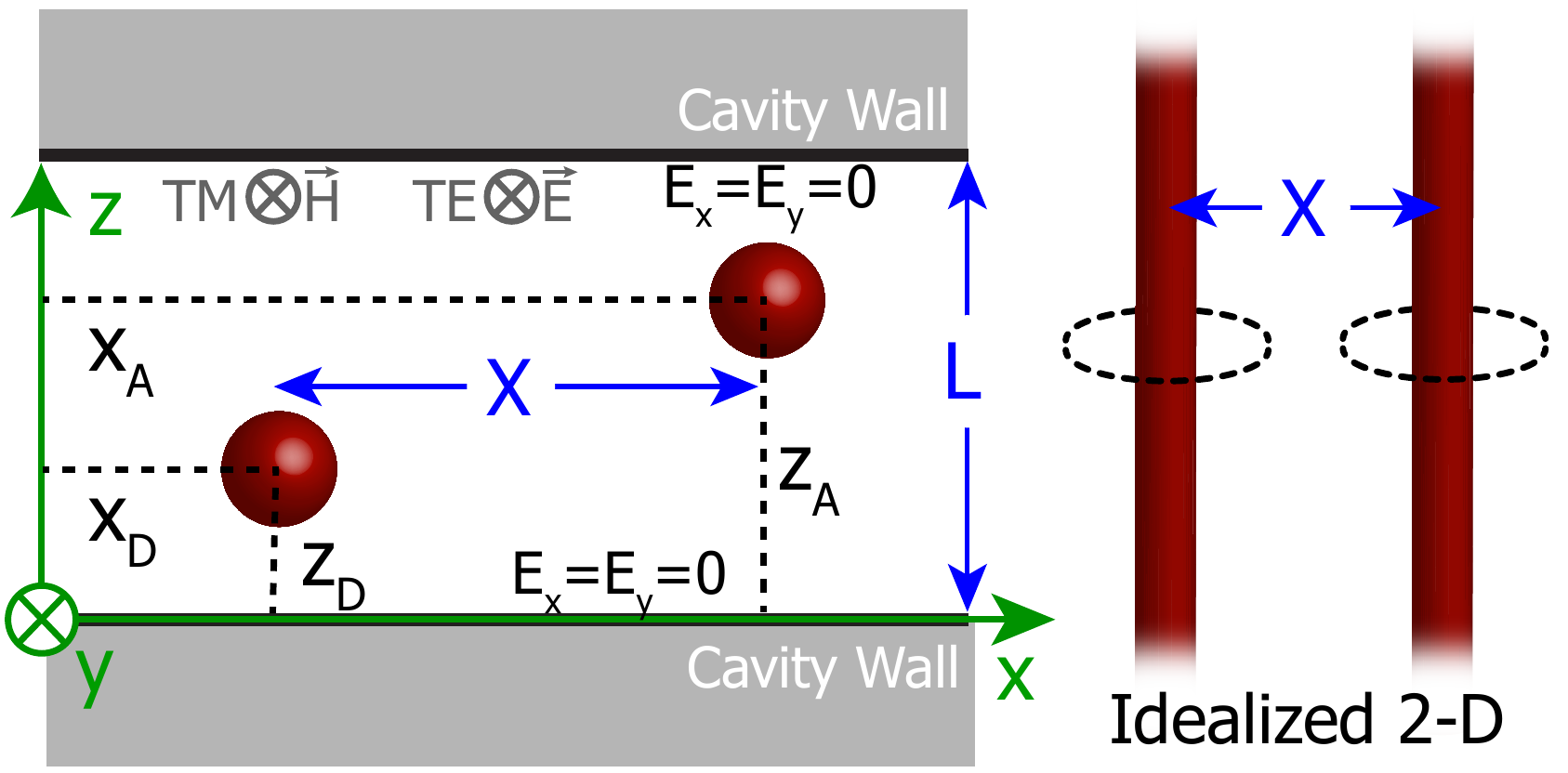}}\\
\subfigure[ ]{\centering{}\includegraphics[width=0.50\textwidth]{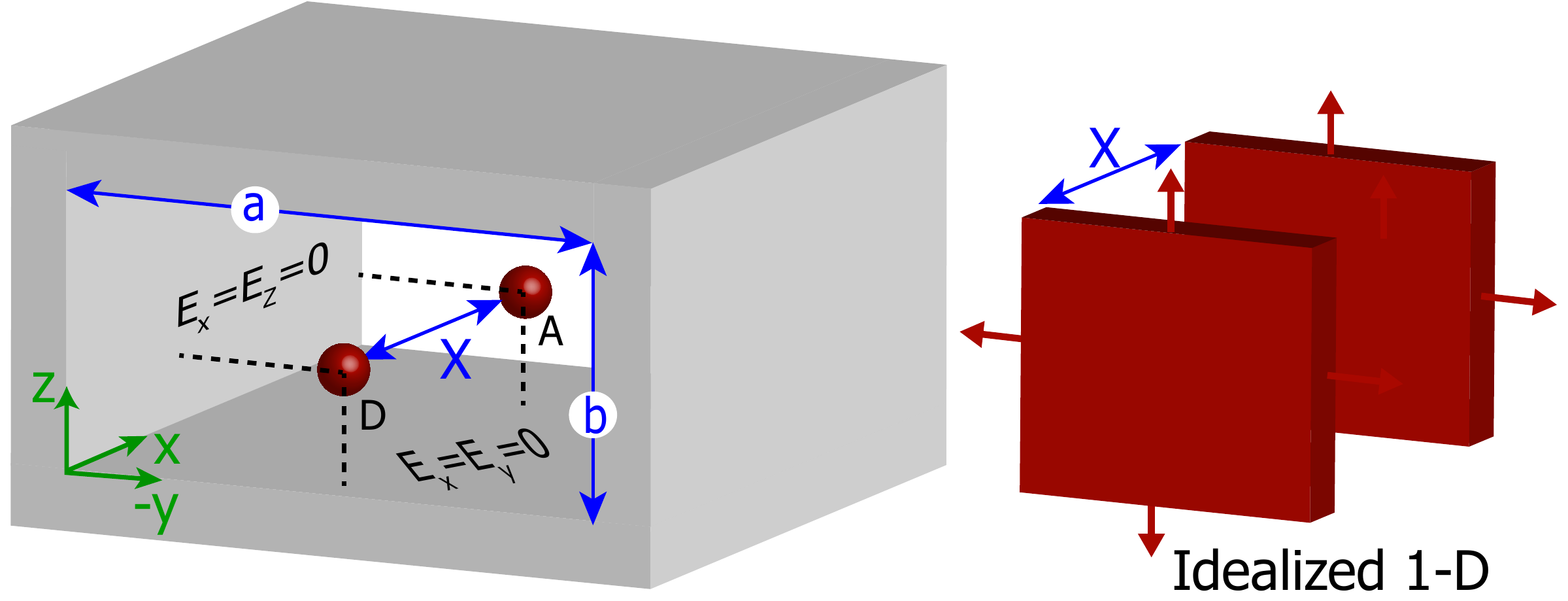}}\\
\caption{\emph{Geometries Considered.} Planar (top) and Rectangular Channel (bottom) cavity geometries along with there reduced dimensional counterparts at right. The donor (D) and acceptor (A) species are depicted as small red spheres within the cavities.}
\label{Geometry}
\end{figure}

Optical cavities may also reduce the extreme near-field sensitivity to donor/acceptor separation, R, which is proportional to $R^{-6}$ for RET and ETU and $R^{-12}$ for EP. Within the abstraction of reduced dimensionality, for instance, this is certainly the case~\cite{Ganainy_NJP_2013, Andrews_2015}. Two-dimensional RET has an $~R^{-4}$ near-field fall off with separation, while its one-dimensional counterpart is not sensitive to separation at all. Along similar lines, it has been found that RET between quantum disks in a quasi-two-dimensional setting has an effective $R^{-1}$ sensitivity to separation~\cite{Parascandolo_2005}. Each geometric setting is depicted in Fig. \ref{Geometry}.

There is a fundamental difference between such reduced dimensional dynamics and those that occur in a cavity though. This is the result of a substantial reduction in what will be referred to here as \emph{Radiation-Dominant} (RD) electromagnetic modes. From a classical electromagnetics perspective, this was noted over twenty years ago~\cite{Kobayashi_PRA_1995} as an effect attributable to evanescent cavity modes. Separation sensitivity was not explored, though, and a mistake in carrying out the complex contour integration resulted in incorrect expressions for the dipole coupling tensor. Cavity RET was subsequently considered between idealized continua of donor and acceptor layers, but separation sensitivity was not taken up\cite{Basko_2000}. In a more recent computational analyses of cavity RET for quantum disks, the ansatz did not account for the cavity modes responsible for the difference in separation sensitivity as compared to that of reduced dimensional idealizations~\cite{Piermarocchi_PRB_2011}. 

This has motivated us to quantify the donor/acceptor separation sensitivity of resonant energy transfer and exciton upconversions within optical cavities. This is accomplished by deriving the Electric Dipole-Dipole Coupling Tensor for both planar and channel cavities. Its components are then analyzed as a function of the relative positions of the donor and acceptor, both within a cavity cross-section and along the cavity axes. The results show a clear distinction between processes in reduced dimensions and their counterparts in optical cavities---i.e. two wires in free space versus two point dipoles in a planar cavity, and two layers in free space versus two point dipoles in a channel cavity. The coupling expressions derived are intended to be helpful in designing systems that optimize the efficiency of desired excitonic processes relative to other relaxation channels that may be available. They are also relevant in the exploration of RET, ETU and EP analogs to Dicke superradiance in cavities~\cite{Mlynek_2014} and as a reference point when considering excitonic dynamics within the strong coupling regime~\cite{Ganainy_NJP_2013}. The methodology developed can be used to quantify exciton dynamics in plasmonic grating settings as well.

%
\section{METHODOLOGY}

Provided that the coupling is sufficiently weak, light/matter interactions between a donor and acceptor can be considered within a perturbative Quantum Electrodynamics (QED) framework\cite{Andrews_2004}. We suppose that the base Hamiltonian, $\hat{H}_0$, is comprised of independent light and matter contributions and restrict attention to the idealization that excitons can be treated as indivisible particles:
\begin{equation}
\hat{H}_0  = \hat{H}_{\mathrm {ex}} +  \hat{H}_{\mathrm {light}} , \quad
\hat{H}_{\mathrm {ex}}  = \sum_{j}\varepsilon_{j}\hat{c}^{\dagger}_{j}\hat{c}_{j} . 
\label{Hcomponents}
\end{equation}
The purely excitonic component, $\hat{H}_{\mathrm {ex}}$, is in terms of the exciton annihilation operator, $\hat{c}_{j}$, of material state $\ket{j}_{\mathrm {ex}}$ with bosonic commutation relations $[\hat{c}_{i}, \hat{c}^{\dagger}_{j}]_-=\delta_{ij}$. The photon component, $\hat{H}_{\mathrm {light}}$, can be expressed in terms of its own annihilation operator once the  electromagnetic modes are identified, and this will now be taken up.

For a prescribed cavity geometry, the normalized electric and magnetic eigenmodes, $\boldsymbol{\varphi}$ and $\boldsymbol{\chi}$, are obtained by solving eigenvalue problems derived from Maxwell's equations,
\begin{equation}
\nabla^2 \boldsymbol{\varphi}^{\mathbf{(\lambda, k)}} = -k^2 \boldsymbol{\varphi}^{\mathbf{(\lambda, k)}}, \quad
\nabla^2\boldsymbol{\chi}^{\mathbf{(\lambda, k)}} = -k^2 \boldsymbol{\chi}^{\mathbf{(\lambda, k)}},
\label{EVP}
\end{equation}
and then applying the requisite boundary conditions. The modes are parametrized by vector $\bf k$ and polarizations $\lambda = 1, 2$. For the sake of clarity, the permittivity, $\varepsilon$, and permeability, $\mu$, have been assumed to be constant. The speed of light in the cavity is denoted by $c$ and $k\equiv|\mathbf{k}|$. 

The electric and magnetic field operators can then be constructed in terms of these modes:
\begin{equation}
\mathbf{\hat{E}}(\mathbf{r}, t)=\sum_{\mathbf{k}}\mathbf{\hat{E}}_{\mathbf{k}}(\mathbf{r}, t) = \sum_{\lambda,\mathbf{k}}E_0(k)\imath \hat{a}^{(\lambda, \mathbf{k})}\boldsymbol{\varphi}^{(\lambda, \mathbf{k})}(\mathbf{r}){\mathrm e}^{\imath c k t}+ \rm{H.c.} \label{eq:Efield}
\end{equation}
\begin{equation}
\mathbf{\hat{H}}(\mathbf{r}, t)=\sum_{\mathbf{k}}\mathbf{\hat{H}}_{\mathbf{k}}(\mathbf{r}, t)=\frac{1}{c \mu}\sum_{\lambda,\mathbf{k}}E_0(k)\imath\hat{a}^{(\lambda, \mathbf{k})}\boldsymbol{\chi}^{(\lambda, \mathbf{k})}(\mathbf{r}){\mathrm e}^{\imath c k t} + \rm{H.c.} \label{eq:Hfield}
\end{equation}
The annihilation operator, $\hat{a}^{\mathbf{(\lambda, k)}}$, destroys a photon in orientation and mode $(\lambda,\mathbf{k})$ and obeys the bosonic commutation relations\cite{Andrews_SPIE_2013}:
\begin{equation}
[\hat{a}^{\mathbf{(\lambda, k)}}, \hat{a}^{(\gamma,\mathbf{p})}]_- = (8 \pi^3 \Omega)^{-1}\delta(\mathbf{k} -  \mathbf{p})\delta_{\lambda, \gamma} .
\end{equation}
where $\Omega$ is the quantization volume. For each cavity, the quantization function, $E_0(k)$, is determined from the following energy relation:
\begin{equation}
\hbar ck=\frac{1}{2}\left(\varepsilon |\langle\mathbf{\hat{E}}_{\mathbf{k}}\rangle|^2+\mu|\langle\mathbf{\hat{H}}_{\mathbf{k}}\rangle|^2\right) =\frac{1}{2}\varepsilon E_0(k)^2 \sum_{\lambda}\int_{\Omega} \left|\boldsymbol{\varphi}^{\mathbf{(\lambda, k)}}(\mathbf{r})\right|^2+\left|\boldsymbol{\chi}^{\mathbf{(\lambda, k)}}(\mathbf{r})\right|^2 d^3\mathbf{r} . \label{eq:E0}
\end{equation}
The light Hamiltonian of Eq. \ref{Hcomponents} is then just $\hat{H}_{\mathrm {light}} =  \sum_{\lambda,\mathbf{k}} \hbar c k \,\hat{a}^{(\lambda, \mathbf{k})\dagger} \hat{a}^{(\lambda, \mathbf{k})}$.

The perturbative light-matter interaction, ${\hat H}_1$, may now be introduced as $\hat{H}_1 = -\boldsymbol{\hat{\mu}}\cdot \mathbf{\hat{E}}$, where $\boldsymbol{\hat{\mu}} = e \hat{\bf{r}}$ acts on the excitonic states, $e$ the magnitude of charge, $\hat{\bf{r}}$ is the exciton position operator, and $\mathbf{\hat{E}}$ acts on the optical states. A more complete accounting of exciton relaxation channels would explicitly account for exciton-phonon interactions in the Hamiltonian~\cite{Piermarocchi_PRB_2011, Lusk_PRB_2015}, but phonon effects can also be  implicitly considered  by adding a broadening term to the eigenvalues of the purely excitonic Hamiltonian. An analogous broadening approach, applied to the optical modes, can be used to account for photon and phonon losses in an imperfect cavity. We considered the broadening  approach, but this causes a decay in the far field and so does not allow a direct comparison with previous low-dimensional results in that regime~\cite{Andrews_2015}. A third approach was therefore adopted in which the cavity was excited just off resonance. In the near field, the resulting dipole coupling tensor was found to match that of the resonant cavity with broadening provided the broadening was of the same order as the detuning of the cavity. 

The initial (i) and final (f) eigenstates are represented in an occupation formalism as state vectors $\ket{i} = \ket{1,0; 0}$ and $\ket{f} = \ket{0,1; 0}$. The first two arguments are for the donor (D) and acceptor (A), respectively, while the third component gives the number of photons, $j$, occupying a given mode and polarization, $j^{(\lambda, \mathbf{k})}$. The photon occupation allows an explicit accounting of all possible intermediate states through which the system may pass, with all paths contributing to the dipole coupling tensor.

The rate analysis is significantly simplified by making a point dipole approximation---i.e. by assuming that the dipole moment of the donor or acceptor is located at its geometric center. Then operator $\mathbf{\hat{r}}$ can be replaced with the position vectors of each species,  $\bf{r}_D$ and $\bf{r}_A$, and the interaction Hamiltonian can be expressed in terms of the transition dipoles of each species, $\boldsymbol{\mu}^{\mathrm{(D)}}$ and $\boldsymbol{\mu}^{\mathrm{(A)}}$:
\begin{equation}
\hat{H}_1 = -\boldsymbol{\mu}^{\mathrm{(D)}}\cdot \mathbf{\hat{E}}({\bf r}_D, t) - \boldsymbol{\mu}^{\mathrm{(A)}}\cdot \mathbf{\hat{E}}({\bf r}_A, t) .
\end{equation}

Since the electric field operator changes the photon occupation, second order perturbation is required to describe resonant energy transfer\cite{Jenkins2003}. The transition rate is thus given by
\begin{equation}
\Gamma=\frac{2\pi}{\hbar}\left|M\right|^{2}\delta(\varepsilon_{f} - \varepsilon_{i})\label{eq:FGR}
\end{equation}
where
\begin{equation}
M=\Bra{f} \sum_{m}\frac{\hat{H}_1\Ket{m}\Bra{m}\hat{H}_1}{\varepsilon_{i} - \varepsilon_{m}} \ket{i}\label{eq:Mfi}
\end{equation}
is the transition amplitude expressed in terms a sum over mediating virtual states, $\Ket{m}$. This can be written out explicitly for a prescribed exciton energy, $\hbar c p$~\cite{Jenkins2003}:
\begin{equation}
M(\mathbf{p},\mathbf{r}_A,\mathbf{r}_D) = \sum_{i,j}\mu_i^{(A)}V_{ij}(\mathbf{p},\mathbf{r}_A,\mathbf{r}_D)\mu_{j}^{(D)} .\label{eq:Mfi2}
\end{equation}
Note that no plane wave assumption has been made regarding the electric field modes; the parametrization of energy using the scalar, $p$, is for analytical convenience only.

The central quantity in such rate formulations is clearly V$_{ij}$, the Cartesian projections of the \emph{Electric Dipole Coupling Tensor}, which is constructed using two time orderings:
\begin{equation}
V_{ij}(\mathbf{p},\mathbf{r}_A,\mathbf{r}_D)=V_{ij}^{+}(\mathbf{p},\mathbf{r}_A,\mathbf{r}_D)+V_{ij}^{-}(\mathbf{p},\mathbf{r}_A,\mathbf{r}_D)\label{eq:Vij}
\end{equation}
where
\begin{equation}
V_{ij}^{+}(\mathbf{p},\mathbf{r}_A,\mathbf{r}_D)=\sum_{\lambda,\mathbf{k}}\frac{E_0(k)^2}{\hbar c}\left(\frac{\varphi_{i}^{(\lambda,\mathbf{k})}(\mathbf{r}_A)\varphi_{j}^{(\lambda,\mathbf{k})*}(\mathbf{r}_D)}{p-k}\right)\label{eq:Vijplus}
\end{equation}
and
\begin{equation}
V_{ij}^{-}(\mathbf{p},\mathbf{r}_A,\mathbf{r}_D)=\sum_{\lambda,\mathbf{k}}\frac{E_0(k)^2}{\hbar c}\left(\frac{\varphi_{i}^{(\lambda,\mathbf{k})*}(\mathbf{r}_A)\varphi_{j}^{(\lambda,\mathbf{k})}(\mathbf{r}_D)}{-p-k}\right) .\label{eq:Vijminus}
\end{equation}
%
%
The $V_{ij}^{+}$ component may be interpreted as being associated with the emission of a virtual photon at the donor which is then absorbed by the acceptor. Its counterpart, $V_{ij}^{-}$, accounts for virtual photons emitted from the donor that travel backwards in time to the acceptor. Both are subject to time-energy uncertainty, of course, but the second path, which can also be thought of as a brief borrowing of energy from the vacuum, would not be possible in a classical enforcement of conservation of energy.

RET, ETU and EP are all based on the same dipole coupling tensor, and donor-acceptor separation sensitivity dwells exclusively there for each process\cite{Andrews_2004, Lacount_2015}. For the RET and ETU, the rate is proportional to the square of the coupling tensor while the EP rate is proportional to its fourth power. 

\section{RESULTS}
\subsection{Free Space}
The dipole coupling tensor is first presented in the standard free-space setting to introduce notation and facilitate a direct comparison with its counterpart in two cavity settings. In free space, the electromagnetic eigenmodes are just plane waves of course,
\begin{eqnarray}
\boldsymbol{\varphi}^{(\lambda,\mathbf{k})}(\mathbf{r}) &=& \hat{\mathrm{\mathbf e}}^{(\lambda,\mathbf{k})} e^{i\mathbf{k}\cdot\mathbf{r}} \\ \nonumber
\boldsymbol{\chi}^{(\lambda,\mathbf{k})}(\mathbf{r}) &=& (\hat{\textbf{k}}\times\hat{\mathrm{\mathbf e}}^{(\lambda,\mathbf{k})})e^{i\mathbf{k}\cdot\mathbf{r}} , \label{eq:Free}
\end{eqnarray}
where $\hat{\bf k} = {\bf k}/k$ and orthonormal polarization vectors, $\hat{\mathrm{\mathbf e}}^{(\lambda,\mathbf{k})}$, are such that $\hat{\bf k} \cdot \hat{\mathrm{\mathbf e}}^{(\lambda,\mathbf{k})} = 0$.  Eq. \eqref{eq:E0} then delivers the standard quantization factor of $E_0(k)=\sqrt{\frac{\hbar ck}{2 \Omega \varepsilon}}$.

\begin{figure}[t]\begin{center}
\includegraphics[width=0.70\textwidth]{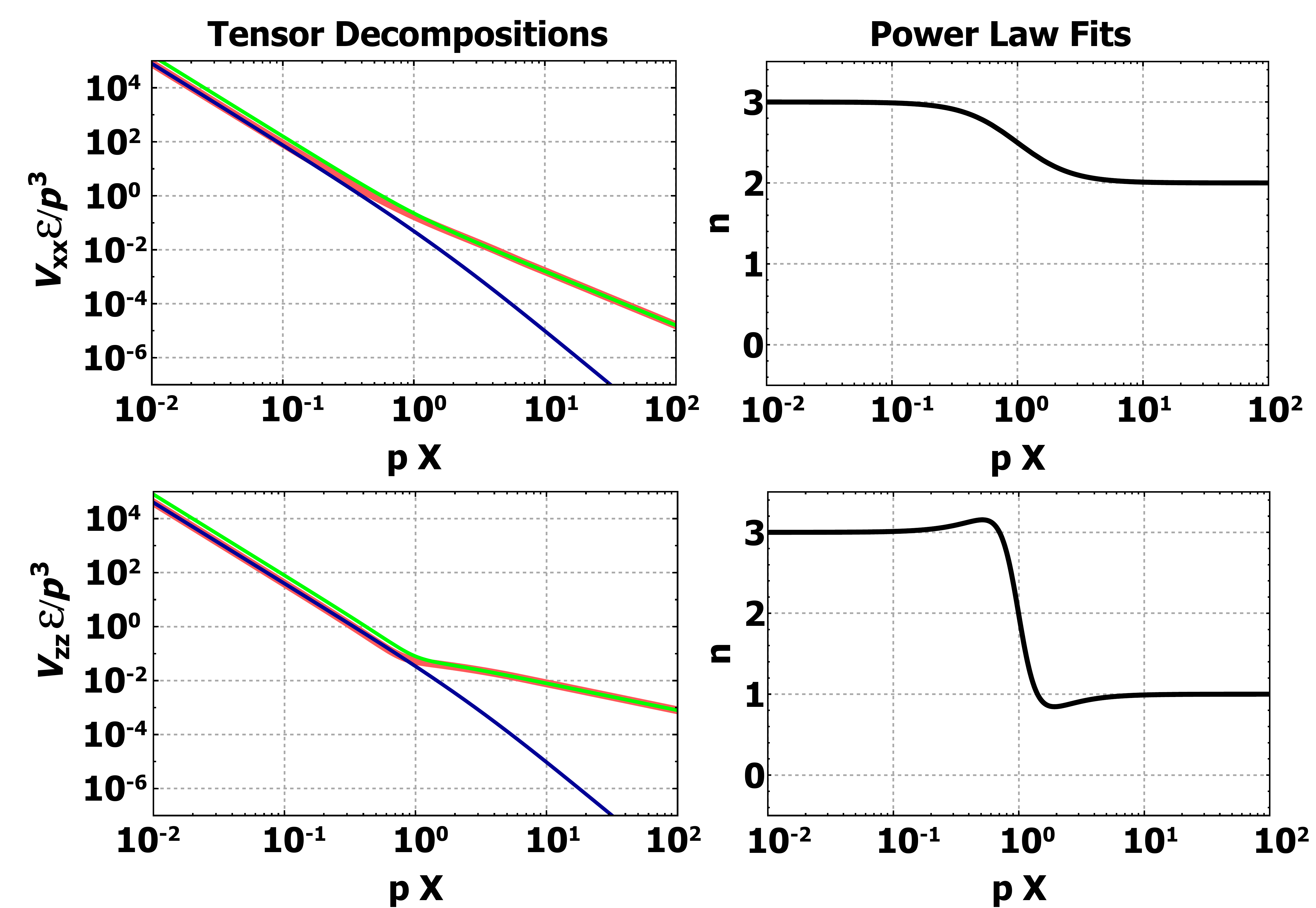}
\caption{\emph{Free-Space Dipole Coupling.} Comparison of the contributions of $V_{ij}^+$ (red/gray), $V_{ij}^-$ (blue/dark gray) and their sum (green/light gray) for $|V_{xx}|$  and $|V_{zz}|$  along with their power-law dependencies in the panels at right.}
\label{V_Free}
\end{center}
\end{figure}

The modes and quantization factor are applied to Eq. \ref{eq:Vij}, the sum over $\bf{k}$ is expressed as an integral which is then evaluated using complex contour integration, and the resulting components of the dipole coupling tensor are found to be \cite{Andrews_2004}:
\begin{equation}
V_{ij}^{\pm}(p,\mathbf{r}) = \frac{1}{4\pi^2\varepsilon r^3}\biggl((\delta_{ij} - \hat{r}_i\hat{r}_j)\Lambda^{\pm}(p r)+(\delta_{ij} - 3\hat{r}_i\hat{r}_j)\Xi^{\pm}(p r)\biggr) .
\label{eq:VijpmFree}
\end{equation}
Here $\mathbf{r} = \mathbf{r}_A - \mathbf{r}_D$ is the position vector pointing from donor to acceptor. In addition, functions $\Lambda^{\pm}(p r)$ and $\Xi^{\pm}(p r)$ are:
\begin{eqnarray}
\Lambda^{\pm}(p r) &=& \mp p r+k^2r^2({\rm cos}(p r){\rm si}(\mp p r)\pm {\rm sin}(p r){\rm Ci}(\mp p r)) \\ \label{eq:Lambda_Xi}
\Xi^{\pm}(p r) &=& -{\rm cos}(p r){\rm si}(\mp p r)\mp {\rm sin}(p r){\rm Ci}(\mp p r)-p r({\rm sin}(p r){\rm si}(\mp p r)\mp {\rm cos}(p r){\rm Ci}(\mp p r)) . \nonumber
\end{eqnarray}
Scalars $p$ and $r$ are the magnitudes of the $\mathbf{p}$ and $\mathbf{r}$ vectors. Functions ${\rm Ci}(p r)$ and ${\rm si}(p r)$ are the cosine integral and shifted sine integral. Without loss of generality, assume that the separation of the donor and acceptor is in the x-direction and let $X \equiv r$. The components of the coupling tensor are plotted in Fig. \ref{V_Free}, where the $\pm$ convention is consistent with Eqs. (\ref{eq:Vij}-\ref{eq:Vijminus}) where "+" signifies a virtual photon being emitted by the donor while  "-" indicates emission by the acceptor. The problem symmetry implies that $|V_{yy}| = |V_{zz}|$ and, because the donor and acceptor both lie on the y-axis, the off-diagonal elements of $V_{ij}$ are zero. In the near field ($p X\ll 1$), both time ordering contributions are of the same size, but in the far field ($p X\gg 1$) the donor emission contribution is much larger than the acceptor emission contribution. The decline of the latter in the far field is due to decreasing energy uncertainty with increasing photon propagation time\cite{Jenkins2body}.

The character of the coupling components is easily distilled to a power-law dependence on the separation distance, $X$, and the fitted exponent, $n$, is plotted for $V_{xx}$ and $V_{zz}$ in Fig. \ref{V_Free}. As expected, the coupling decays as $1/X^3$ in the near field for nonradiative energy transfer. In the far field, the dependence changes to $1/X^2$ for $V_{xx}$ and $1/X$ for $V_{zz}$---radiative energy transfer. The transition from near-field to far-field behavior occurs at $p X\sim 1$, and this boundary is best explained by first expressing the dipole coupling in a spherical-wave expansion to make it reflect free spatial symmetry. The separation dependence is then described~\cite{Cappellin_2008} by $n^{\rm{th}}$-order spherical Hankel functions of the first kind, and these have a transition region at $p X \sim n$. This also indicates that the energy transfer in the far field is dominated by the lowest order spherical Hankel function.

%

%
\subsection{PLANAR CAVITY}
We now turn attention to the primary focus of this paper, quantifying the separation sensitivity of cavity-based RET, ETU and EP. First consider the planar cavity Fig. \ref{Geometry}(a) consisting of two perfectly conducting plates located at $z = 0$ and $z = L$. These mirrors impose boundary conditions that must be satisfied by the eigenmodes of Eq. \ref{EVP}. Continuity of the electric field at the boundaries implies that the tangential components of the electric field need to be zero there. Along the same lines, the normal component of the magnetic field is zero at the boundaries. The requisite electric modes are then
\begin{eqnarray}
\boldsymbol{\varphi}^{(TE,\mathbf{k})}(\mathbf{r}) &=& {\rm sin}(k_z z) e^{i\mathbf{k_\gamma}\cdot\mathbf{r}}(\hat{z}\times\hat{k}_{\gamma})\\ \nonumber
\boldsymbol{\varphi}^{(TM,\mathbf{k})}(\mathbf{r})  &=& \frac{i k_z}{k}\text{sin}(k_z z) e^{i\mathbf{k_\gamma}\cdot\mathbf{r}}\hat{k}_{\gamma}-\frac{k_{\gamma}}{k}{\rm cos}(k_z z) e^{i\mathbf{k_\gamma}\cdot\mathbf{r}}\hat{z}\label{eq:phi_Planar}
\end{eqnarray}
while the magnetic modes are 
\begin{eqnarray}
\boldsymbol{\chi}^{(TE,\mathbf{k})}(\mathbf{r}) &=& {\rm cos}(k_z z) e^{i\mathbf{k_\gamma}\cdot\mathbf{r}}(\hat{z}\times\hat{k}_{\gamma}) \\ \nonumber
\boldsymbol{\chi}^{(TM,\mathbf{k})}(\mathbf{r}) &=& \frac{i k_z}{k}{\rm cos}(k_z z) e^{i\mathbf{k_\gamma}\cdot\mathbf{r}}\hat{k}_{\gamma}+\frac{k_{\gamma}}{k}{\rm sin}(k_z z) e^{i\mathbf{k_\gamma}\cdot\mathbf{r}}\hat{z} . \label{eq:chi_Planar}
\end{eqnarray}
The boundary conditions imply that the z-component of the wavenumber is discrete, $k_z = n\pi/L$, making it natural to define two-dimensional vector, $\bf{k}_{\gamma}$, in the x,y-plane. As usual, mode polarizations, $\lambda$, are referred to as transverse electric (TE) and transverse magnetic (TM), where transverse implies that there is no field component perpendicular to the boundary. The quantization factor associated with the planar cavity is, from Eq. \eqref{eq:E0}, $E_0(k)=\sqrt{\frac{\hbar ck}{\Omega \varepsilon}}$.

These preliminary quantities are sufficient to now derive the dipole coupling tensor using Eqs. (\ref{eq:Vij} - \ref{eq:Vijminus}). Construction of its V$_{xx}$ projection is detailed in Appendix A, while analogous work for the other components can be found in the Supporting Information. The final expressions are:
\begin{multline}
V_{xx}(p,X,z_A,z_D) = \frac{-i}{4\varepsilon L}\sum_{k_z} {\rm sin}(k_z z_A) {\rm sin}(k_z z_D)\\
\left( \left(p^2+k_z^2\right) H_0^{(1)}\left( X\sqrt{p^2-k_z^2}\right)+\left( p^2-k_z^2\right)H_2^{(1)}\left(X\sqrt{p^2-k_z^2}\right) \right)
\label{eq:Vxx8Planar}
\end{multline}
\begin{multline}
V_{yy}(p,X,z_A,z_D) = \frac{-i}{4\varepsilon L}\sum_{k_z} {\rm sin}(k_z z_A) {\rm sin}(k_z z_D)\\
\left( \left(p^2+k_z^2\right) H_0^{(1)}\left( X\sqrt{p^2-k_z^2}\right)-\left( p^2-k_z^2\right)H_2^{(1)}\left(X\sqrt{p^2-k_z^2}\right) \right)
\label{eq:VyyPlanar}
\end{multline}
\begin{equation}
V_{zz}(p,X,z_A,z_D) = \frac{-i}{2\varepsilon L}\sum_{k_z} {\rm cos}(k_z z_A) {\rm cos}(k_z z_D)
\left(p^2-k_z^2\right)\left(H_0^{(1)}\left( X\sqrt{p^2-k_z^2}\right)\right)
\label{eq:Vxx8Planar}
\end{equation}
\begin{multline}
V_{xz}(p,X,z_A,z_D) = \frac{-i}{4\varepsilon L}\sum_{k_z} {\rm sin}(k_z z_A) {\rm cos}(k_z z_D)X k_z(p^2-k_z^2)\\
\left(H_0^{(1)}\left( X\sqrt{p^2-k_z^2}\right)+H_2^{(1)}\left(X\sqrt{p^2-k_z^2}\right) \right) .
\label{eq:VxzPlanar}
\end{multline}
Because we have chosen to have no separation in the y-direction, $V_{xy} = V_{yz} = 0$. If there is no separation in the z-direction, then $V_{xz} = 0$ as well. The sums in the $V_{ij}$ expressions can be partitioned into two distinct sets of terms. The set for which $p^2 - k_z^2 < 0$ is associated with a state sequence in which the total energy of the system during the energy transfer is higher than the exciton energy. This includes only states in which virtual photons have an energy greater than the exciton energy.  These will be referred to as \emph{Non-Radiation-Dominant} (NRD) terms since they decay in the far field as the time-energy uncertainty inequality narrows the range of allowed energies. The set for which $p^2 - k_z^2 > 0$  is associated with a state sequence in which the total energy of the system during the energy transfer \emph{can} be equal to the exciton energy. This includes all states in which virtual photons have an energy equal to the exciton energy and both donor and acceptor are in their ground states. As separation distance increases, time-energy uncertainty causes this set to narrow towards that of energy conserving photon emission and absorption, so they are referred to as \emph{Radiation Dominant} contributions. It is worth emphasizing that the Radiation Dominant set still includes quantum pathways that borrow energy from the vacuum and influence near-field dynamics. Unlike the case in free space, the overall expression does not lend itself to a simple analytic decomposition into forward and backward propagating photon contributions. While this could be accomplished numerically, that was not carried out in order to have analytic results.

The planar cavity dipole coupling tensor can be compared directly with its counterpart derived within a two-dimensional setting to address the primary focus of this investigation---i.e. the difference in dynamics due to cavity constraints as opposed to a simple reduction in dimensionality. The two-dimensional dipole coupling tensor has been previously found to be~\cite{Andrews_2015}
\begin{equation}
V_{{\rm 2D}}(p,X) = \frac{-i}{4\varepsilon L} p^2 \left(H_0^{(1)}\left(p X\right)\right) .
\label{eq:VNW}
\end{equation}
Not surprisingly, this expression is essentially the same as the $V_{zz}$ component of the planar cavity coupling tensor under the restriction that $k_z = 0$. With the z-dependence thus removed, $V_{zz}$ is the only non-zero projection of the tensor. The  distinction between the reduced and cavity coupling tensors is then just a factor of two which can be attributed to the difference in quantization factors in free-space and planar cavity settings. 

\begin{figure}[t]\begin{center}
\includegraphics[width=0.70\textwidth]{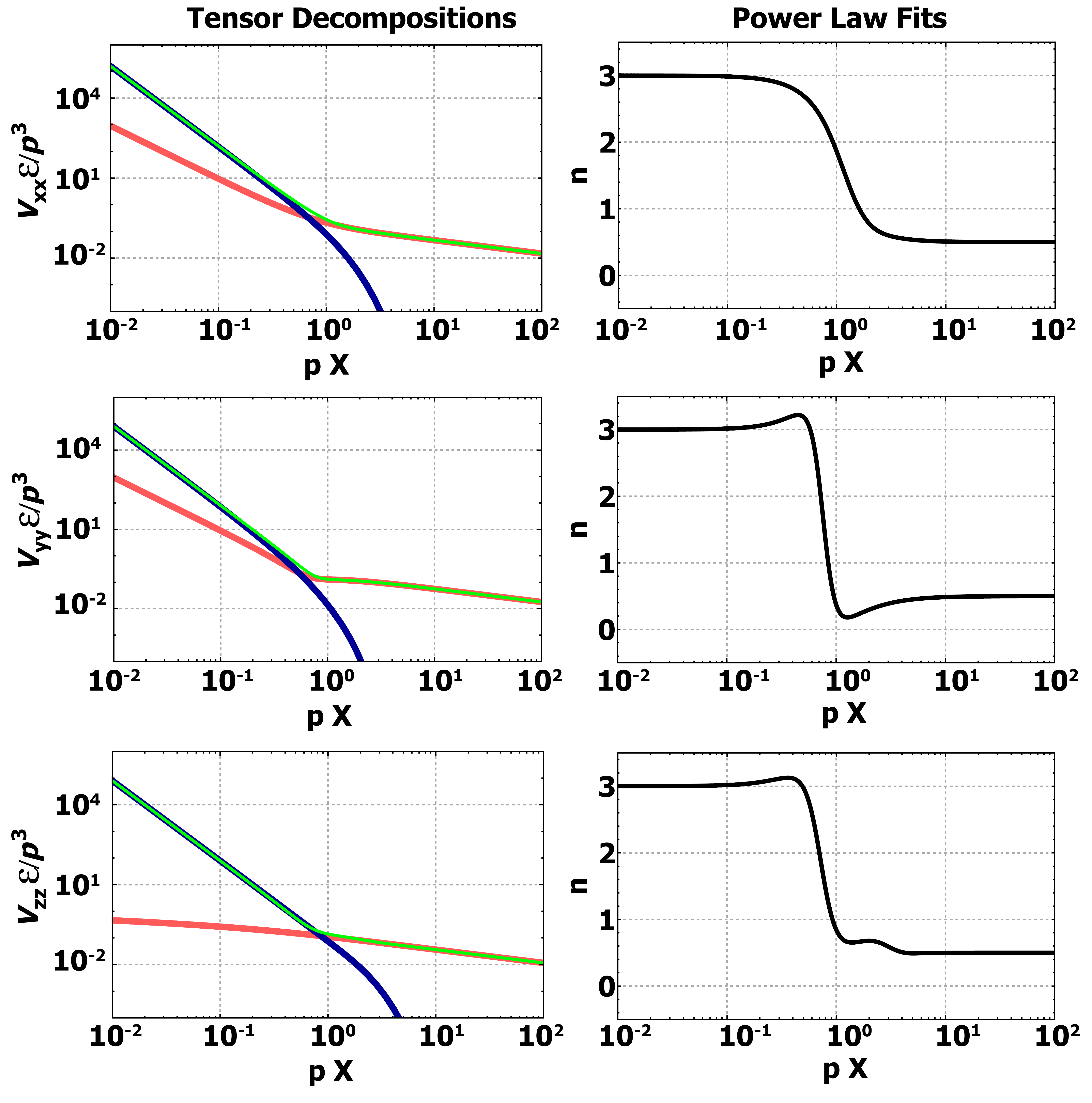}
\caption{\emph{Planar Cavity Dipole Coupling.} Comparison between the RD (red/gray) and NRD (blue/dark gray) contributions to the dipole coupling tensor along with the sums (green/light gray) for $|V_{xx}|$ (a), $|V_{yy}|$ (b) and $|V_{zz}|$ (c). 1/X$^n$ separation dependencies are shown in the right panels.}
\label{V_Planar}
\end{center}
\end{figure}

The full planar cavity coupling tensor has a much richer character, as shown in Fig. \ref{V_Planar}. A cavity width of $L = 1.1\pi/p$---large enough to allow a single non-resonant Radiation Dominant term for each element of the dipole coupling tensor---has been adopted. Moreover, the z-positions of both donor and acceptor are assumed to be halfway between the plates. In the near field ($p r\ll 1$) the NRD terms dominate while the reverse is true in the far field.

Further insight into the role of the cavity can be obtained by fitting the electric dipole coupling to a power-law dependence on separation: $\sim 1/X^n$. The resulting spatially varying functions, $n(X)$,  are plotted in Fig. \ref{V_Planar}. Contrary to previous studies for energy transfer in cavities\cite{Xu_2014}, the near-field coupling decays as $1/X^3$. The earlier work assumed that the RD contribution to $V_{ij}$ was much greater than the NRD component, and the latter was neglected. As is clear from Fig. \ref{V_Planar}, this assumption is only valid in the far field where the separation sensitivity is $~X^{-1/2}$. This makes physical sense because it implies that the far-field energy transfer rate has as a $~1/X$ separation sensitivity as one would expect from two-dimensional wave propagation. The transition from near-field to far-field behavior occurs in the region where $p X \sim 1$.

\begin{figure}[h]
\centering{}
\includegraphics[width=0.7\textwidth]{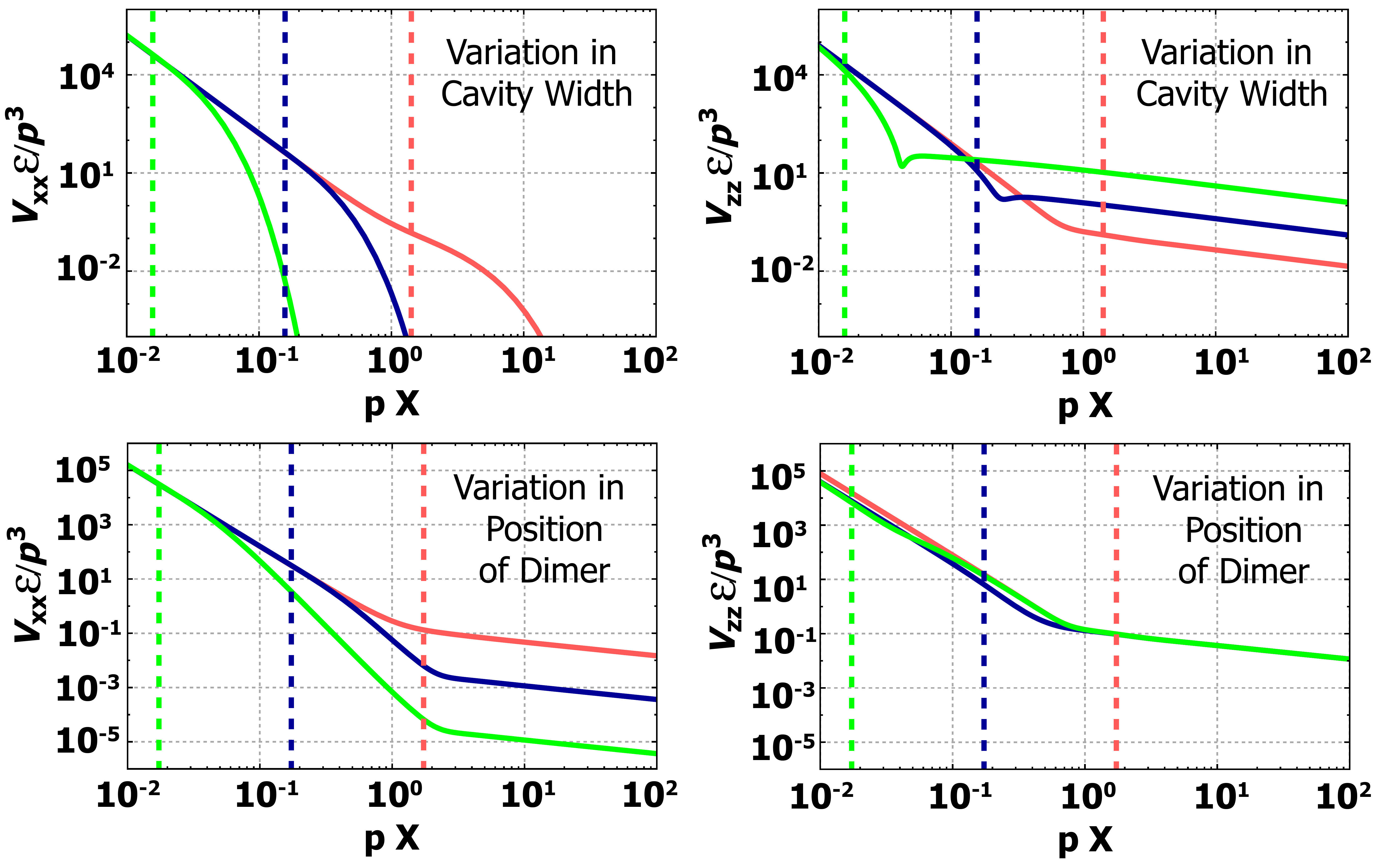}
\caption{\emph{Planar Cavity Geometry Study.} (Top) Influence of planar cavity width on dipole coupling components V$_{xx}$ and V$_{zz}$ for cavities of sizes of: $L = 0.9\pi/p$ (red/gray), $L = 0.1\pi/p$ (blue/dark gray), and $L = 0.01\pi/p$ (green/light gray). Vertical lines are plotted at $X = z = L/2$. (Bottom) Influence of donor and acceptor height within planar cavity on dipole coupling components V$_{xx}$ and V$_{zz}$ for species positioned at $z = L/2$ (red/gray), $z = L/20$ (blue/dark gray), and $z = L/200$ (green/light gray). Vertical lines are plotted at $X = z$.}
\label{Studies_Planar}
\end{figure}

The influence of the geometric setting on the coupling tensor is taken up from two perspectives in Fig. \ref{Studies_Planar}, where the change in its components with cavity size and the effect of moving the donor and acceptor off center are both quantified. A characteristic separation distance is observed to occur at the smaller of $X = z$ and $X = L - z$---i.e. when the separation distance is the same as the distance of donor and acceptor to the nearest wall. This can be explained by imagining the virtual photons as spherical waves propagating outward from one moiety until they reach the other where they are annihilated. The time over which this occurs defines a sphere of influence of radius $X$ for the virtual photons. If the cavity walls do not lie within this sphere, the virtual photon never encounters the cavity boundaries and is not influenced by the constraints they would otherwise impose on it. The virtual photons therefore travel within this sphere just as they would in free space. Additional quantum pathways exist in which the spherical wave is reflected off a cavity wall and then absorbed. However these are necessarily longer than the direct path described previously and so do not significantly contribute to the coupling tensor. This is a fundamental difference separating the physical properties of virtual photons versus real photons within a cavity. A virtual photon has a well-defined creation \emph{and} annihilation point which gives it a finite sphere of influence. Real photons, on the other hand, have either a well-defined creation or annihilation point, but not both, implying an infinite sphere of influence. Real photons are therefore influenced by all boundary conditions while virtual photons are not. This is, of course, a bit circular in the sense that designation as a "real" photon in a cavity implies that it can be described with geometric (ray) optics in its interaction with all cavity boundaries.

%
%
From Fig. \ref{V_Planar} alone, it might be concluded that the dipole coupling of free space and the planar cavity are essentially identical in the near field. While this is true for a relatively large cavities, a substantial deviation develops as the cavity spacing is reduced. Fig. \ref{Studies_Planar} (top right panel) quantifies this. The boundary condition imposed by the cavity walls dramatically decreases $V_{xx}$ (top left panel) and this effect is also manifested in the z-position dependence $V_{xx}$ as well (bottom left panel). Of more technological interest, though, is that $V_{zz}$ exhibits a large increase relative to its free-space counterpart as the cavity narrows.   This is the only setting in which the cavity can be used to increase the rate of RET, ETU and EP. This is observed for $p X > 0.1$. As shown in the bottom right panel, there is only a very weak dependence of $V_{zz}$ on z-position.

\subsection{RECTANGULAR CHANNEL}
Having elucidated the distinction between two-dimensional free space and planar cavities, we now turn attention to rectangular channels as shown in Fig. \ref{Geometry}(b). Four perfectly conducting plates are located at $y = 0$, $y = a$, $z = 0$ and $z = b$, and these mirrors create additional constraints that are satisfied by the following electromagnetic modes:
\begin{eqnarray}
\boldsymbol{\varphi}^{(TE,\mathbf{k})}(\mathbf{r}) &=& \left(\frac{-i k k_z}{k_\eta^2}{\rm cos}(k_y y){\rm sin}(k_z z)\hat{y}+\frac{i k k_y}{k_\eta^2}{\rm sin}(k_y y){\rm cos}(k_z z)\hat{z}\right)e^{i k_x x} \\ \nonumber
\boldsymbol{\varphi}^{(TM,\mathbf{k})}(\mathbf{r}) &=& \left({\rm sin}(k_y y){\rm sin}(k_z z)\hat{x}+\frac{i k_x k_y}{k_\eta^2}{\rm cos}(k_y y){\rm sin}(k_z z)\hat{y}\right. \\ \nonumber
& & \quad\left. - \frac{i k_x k_z}{k_\eta^2}{\rm sin}(k_y y){\rm cos}(k_z z)\hat{z}\right)e^{i k_x x}
\label{eq:phi_Channel}
\end{eqnarray}
and
\begin{eqnarray}
\boldsymbol{\chi}^{(TE,\mathbf{k})}(\mathbf{r}) &=& \left({\rm cos}(k_y y){\rm cos}(k_z z)\hat{x}-\frac{i k_x k_y}{k_\eta^2}{\rm sin}(k_y y){\rm cos}(k_z z)\hat{y}\right. \\ \nonumber
& &\quad - \left.\frac{i k_x k_z}{k_\eta^2}{\rm cos}(k_y y){\rm sin}(k_z z)\hat{z}\right)e^{i k_x x} \\ \nonumber
\boldsymbol{\chi}^{(TM,\mathbf{k})}(\mathbf{r}) &=& \left(\frac{-i k k_z}{k_\eta^2}{\rm sin}(k_y y){\rm cos}(k_z z)\hat{y}+\frac{i k k_y}{k_\eta^2}{\rm cos}(k_y y){\rm sin}(k_z z)\hat{z}\right)e^{i k_x x}.
\label{eq:chi_Channel}
\end{eqnarray}
Here $k_\eta^2 = k_y^2 + k_z^2$. The boundary conditions are satisfied by $k_y = \frac{m\pi}{a}$ and $k_z=\frac{n\pi}{b}$. As with the planar cavity, the polarizations are of either TE or TM character. The quantization factor, from Eq. \eqref{eq:E0}, is found to be
\begin{equation}
E_0(k) = \sqrt{\frac{\hbar ck}{\Omega \varepsilon}\frac{2 k_\eta^2}{k^2}} . \label{eq:E0_Channel}
\end{equation}

This classical electromagnetic setting allows the $V_{xx}$ component of the dipole coupling tensor to be derived (Appendix B), and it is found to be:
\begin{equation}
V_{xx}(p,\mathbf{r}^{(A)},\mathbf{r}^{(D)})=\frac{-2 i}{\varepsilon a b}
\sum_{k_y,k_z}{\rm sin}(k_y y_A) {\rm sin}(k_z z_A){\rm sin}(k_y y_D){\rm sin}(k_z z_D)
\left(\frac{k_\eta^2}{\sqrt{p^2-k_\eta^2}}e^{i X\sqrt{p^2-k_\eta^2}}\right) .
\label{eq:Vxx5WG}
\end{equation}
Similar derivations were used to find the other coupling elements (Supporting Information): 
\begin{equation}
V_{yy}(p,\mathbf{r}^{(A)},\mathbf{r}^{(D)})=\frac{-2 i}{\varepsilon a b}
\sum_{k_y,k_z}{\rm cos}(k_y y_A) {\rm sin}(k_z z_A){\rm cos}(k_y y_D){\rm sin}(k_z z_D)
\left(\frac{p^2-k_y^2}{\sqrt{p^2-k_\eta^2}}e^{i X\sqrt{p^2-k_\eta^2}}\right)
\label{eq:VyyWG}
\end{equation}
\begin{equation}
V_{zz}(p,\mathbf{r}^{(A)},\mathbf{r}^{(D)})=\frac{-2 i}{\varepsilon a b}
\sum_{k_y,k_z}{\rm sin}(k_y y_A) {\rm cos}(k_z z_A){\rm sin}(k_y y_D){\rm cos}(k_z z_D)
\left(\frac{p^2-k_z^2}{\sqrt{p^2-k_\eta^2}}e^{i X\sqrt{p^2-k_\eta^2}}\right)
\label{eq:VzzWG}
\end{equation}
\begin{equation}
V_{xy}(p,\mathbf{r}^{(A)},\mathbf{r}^{(D)})=\frac{-2}{\varepsilon a b}
\sum_{k_y,k_z}{\rm sin}(k_y y_A) {\rm sin}(k_z z_A){\rm cos}(k_y y_D){\rm sin}(k_z z_D)
\left(k_y e^{i X\sqrt{p^2-k_\eta^2}}\right)
\label{eq:VxyWG}
\end{equation}
\begin{equation}
V_{xz}(p,\mathbf{r}^{(A)},\mathbf{r}^{(D)})=\frac{-2}{\varepsilon a b}
\sum_{k_y,k_z}{\rm sin}(k_y y_A) {\rm sin}(k_z z_A){\rm sin}(k_y y_D){\rm cos}(k_z z_D)
\left(k_z e^{i X\sqrt{p^2-k_\eta^2}}\right)
\label{eq:VxyWG}
\end{equation}
\begin{equation}
V_{yz}(p,\mathbf{r}^{(A)},\mathbf{r}^{(D)})=\frac{-2 i}{\varepsilon a b}
\sum_{k_y,k_z}{\rm cos}(k_y y_A) {\rm sin}(k_z z_A){\rm sin}(k_y y_D){\rm cos}(k_z z_D)
\left(\frac{k_y k_z}{\sqrt{p^2-k_\eta^2}}e^{i X\sqrt{p^2-k_\eta^2 }}\right)
\label{eq:VzzWG}
\end{equation}
As with the planar cavity, the sums in the $V_{ij}$ expressions can be partitioned into two types, $p^2-k_\eta^2 < 0$ and $p^2-k_\eta^2 > 0$, which are the RD and NRD contributions, respectively.

The channel components of the dipole coupling tensor can be compared directly to its scalar counterpart derived within a strictly one-dimensional setting~\cite{Andrews_2015}:
\begin{equation}
V_{{\rm 1D}}(p,X)=\frac{-i}{2 a b\varepsilon}\left( p e^{i p X}\right) .
\label{eq:VQW}
\end{equation}

Unlike the planar cavity, we cannot set the cavity modes, $k_\eta$, to zero because then the entire dipole coupling tensor would disappear. However, if we ignore the sine terms we can see that, by setting the cavity modes to zero, the $V_{yy}$ and $V_{zz}$ terms match the reduced dimensions result except for a factor of four. This discrepancy can be traced back to the difference in the quantization factor for the two geometries.

\begin{figure}[t]\begin{center}
\includegraphics[width=0.70\textwidth]{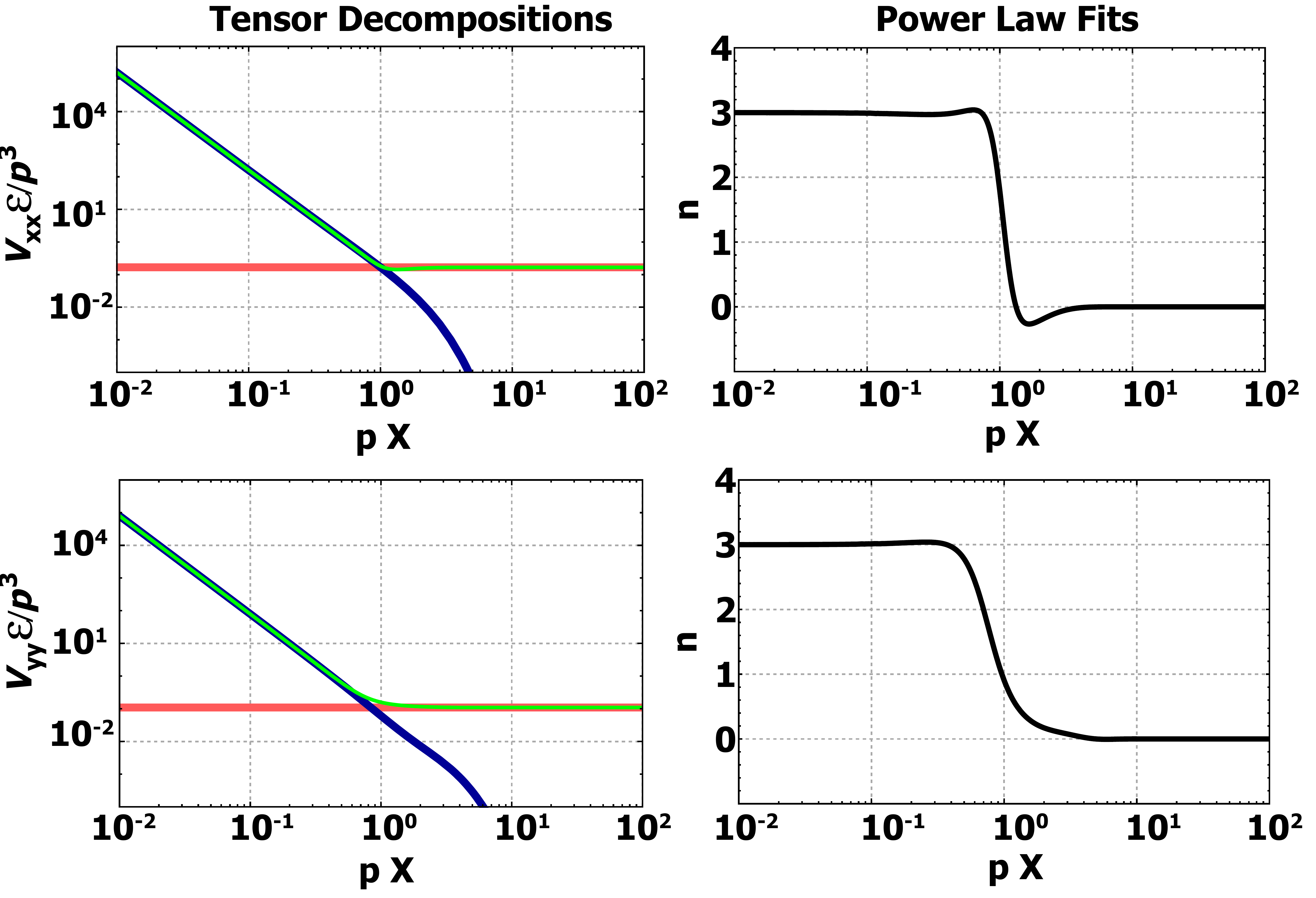}
\caption{\emph{Rectangular Channel Cavity Dipole Coupling.} Plot of the $|V_{xx}|$ and $|V_{yy}|$ components of the dipole coupling tensor between along with their RD (red/gray) and NRD (blue/dark gray) contributions. The 1/X$^n$ separation dependencies are shown in the right panels.}
\label{V_Channel}
\end{center}
\end{figure}

Up to this point, the derivation has been for an arbitrary separation of donor and acceptor within a general rectangular waveguide, but henceforth it is assumed that the waveguide has a square cross-section. Furthermore, unless otherwise stated, the y and z positions of both donor and acceptor are now taken to be equidistant from the waveguide walls. The components of the dipole coupling tensor are plotted in Fig. \ref{V_Channel}. Note that $V_{yy} = V_{zz}$ due to the assumed channel symmetry. We have assumed the waveguide dimensions to be  $a = b = 1.1\sqrt{2}\pi/p$---just large enough to allow a single non-resonant RD term for each element of the dipole coupling tensor. As with the planar cavity it was found that, in the near field ($p X\ll 1$), the NRD terms of $V_{ij}$ are of much greater order than the RD term, but in the far field ($p X\gg 1$) the dominance is reversed.

The character of the coupling components is easily distilled to a power-law dependence on separation distance, $X$, and the fitted exponent, $n$, is plotted for $V_{xx}$ and $V_{yy}$ in Fig. \ref{V_Channel}. Just as in the planar case, the near-field coupling decays as 1/X$^3$ but converges to a constant in the far field. The transition from near-field to far-field behavior occurs in the region where $pX \sim 1$.

\begin{figure}[h]
\centering{}
\includegraphics[width=0.7\textwidth]{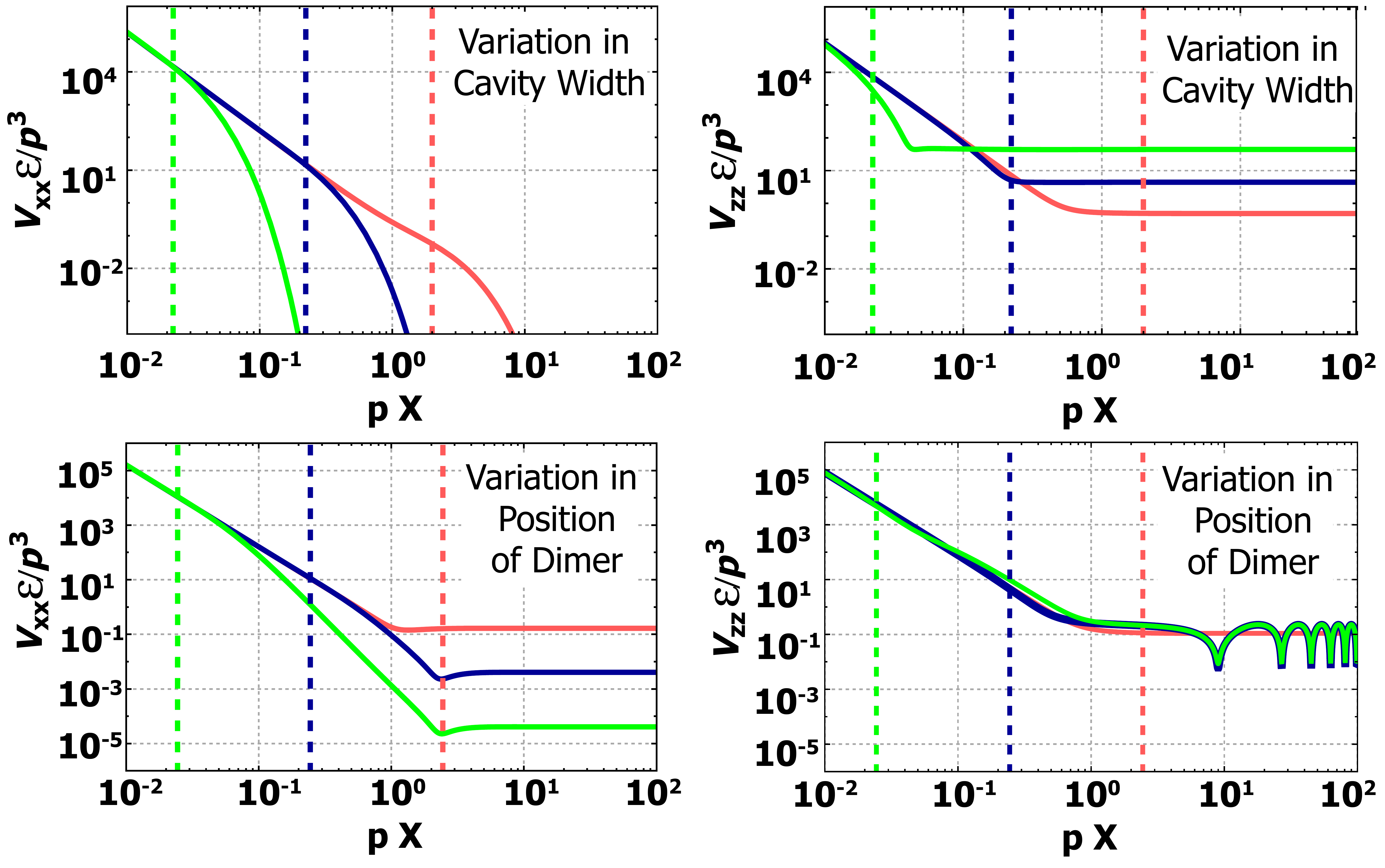}
\caption{\emph{Rectangular Channel Cavity Geometry Study.} (Top) Influence of square channel cavity width on dipole coupling components $V_{xx}$ and $V_{zz}$ for rectangular channel cavities of sizes of $a = 1.1\pi/p$ and $b = 0.9\pi/p$ (red/gray), $L = 0.1\pi/p$ (blue/dark gray), and $L = 0.01\pi/p$ (green/light gray). Vertical lines are plotted at $X = z = L/2$.  (Bottom) Influence of species position within square channel cavity ($a = b = 1.1\sqrt{2}\pi/2$) on dipole coupling component V$_{xx}$ for species positioned at $z = L/2$ (red/gray), $z = L/20$ (blue/dark gray), and $z = L/200$ (green/light gray). Vertical lines are plotted at $X = z$.}
\label{Studies_Channel}
\end{figure}

The role of cavity size and relative donor-acceptor position are summarized in Fig. \ref{Studies_Channel}. Consistent with the planar cavity, a transition in behavior occurs when the nearest cavity wall is at the same distance as the separation between donor and acceptor. So long as the sphere of influence of virtual photons does not reach the channel wall before it is annihilated, the dynamics will be that of free space. Also in line with the planar cavity analysis, narrowing the cavity dimensions can be used to significantly increase the tensor component corresponding to the direction in which the cavity is narrowed. While the near-field z-position dependence was found to be weak, the far-field exhibits an interesting oscillatory behavior when $z \neq L/2$. This is caused by the interaction between the $k_y = \pi/L$ and $k_z = 0,\pi/L$ terms.
%
%
%

%
\section{CONCLUSIONS}

The electric dipole-dipole coupling tensor, $\bf V$, underlies a number of important excitonic processes that include Resonant Energy Transfer (RET), Energy Transfer Upconversion (ETU) and Energy Pooling (EP). Each of these processes is extremely sensitive to the separation distance between donor and acceptor moieties which could be atoms, molecules, quantum dots, or defects in condensed matter. However, the tensor is found to have a much slower decay with separation in theoretically posited, low-dimensional settings. This has generated a discussion of whether or not this sensitivity could also be reduced by encapsulating the active materials within optical cavities. 

\begin{figure}[t]\begin{center}
\includegraphics[width=0.45\textwidth]{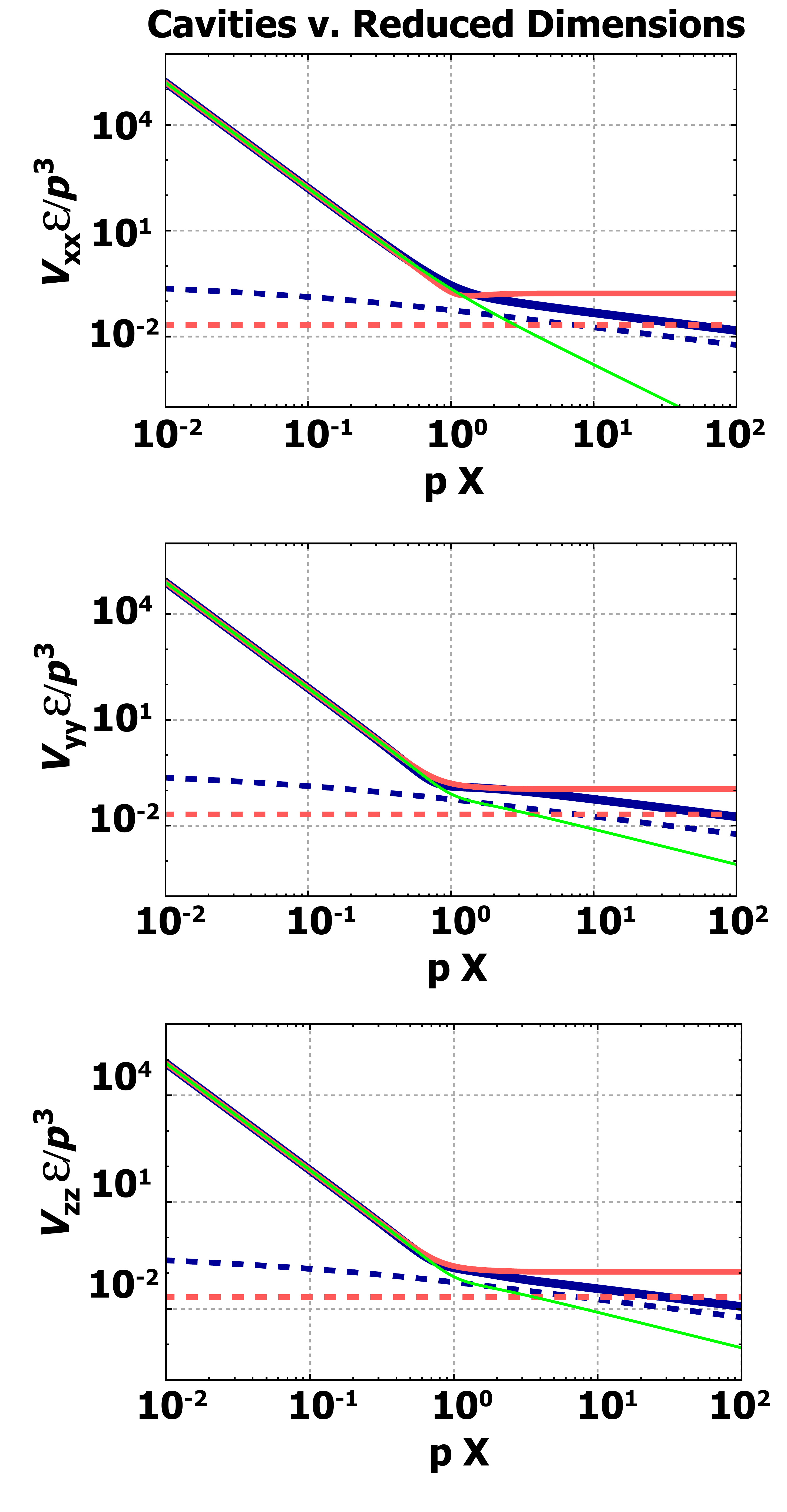}
\caption{\emph{Dipole Coupling Comparison for Free Space, Planar Cavity and Rectangular Channel Cavity.} The diagonal components of the dipole tensor for each geometry: 3-D free space (green/light gray), planar cavity (blue/dark gray), channel cavity (red/gray), 2-D free space (blue/dark gray, dashed), and 1-D free space (red/gray, dashed).}
\label{V_Compare}
\end{center}
\end{figure}

A perturbative, cavity quantum electrodynamics effort was carried out to derive analytical expressions for the coupling tensor components in two types of cavities to facilitate an elucidation of how these tensors behave relative to their free-space counterparts. Key comparisons between five settings can be distilled from the results obtained---the standard three-dimensional free-space coupling; its two lower-dimensional counterparts; and the planar and channel cavity coupling.  For the sake of clarity, the donor and acceptor are positioned in the center of the cavities, a planar cavity wall separation of $1.1\pi/p$ is adopted, and a square channel is considered with a wall separation of of $1.1\sqrt{2}\pi/p$. The results are shown in Fig. \ref{V_Compare}. 

In the near field, where RET, ETU and EP are of primary interest, there is no significant difference between the three-dimensional free-space coupling tensor and its cavity counterparts. Their components, though, are dramatically different than those of the reduced dimensional settings. On the other hand, the far-field coupling terms decay as 1/X$^{(d-1)}$, where $d$ is the number of unconfined dimensions. The more confined the cavity is, the greater the far-field coupling for a given separation between donor and acceptor. It is also worth noting that, in the far field, the magnitude of the cavity couplings are always larger than their reduced-dimensional counterparts. 

The substantial difference between the near-field behavior in cavities and their reduced dimensional counterparts is due to the fact that the cavity boundaries severely limit the number of electromagnetic modes available to virtual photons in mediating an energy transfer. In three-dimensional free space, modes are parameterized by a triad of real numbers while in the planar and channel cavities the parametrization is two (planar) or one (channel) real number along with one (planar) or two (channel) integers. This, in turn, reduces the number of RD photon pathways relative to NRD pathways and has the classical interpretation of evanescent modes dominating propagating modes within cavities.

Although the focus of this investigation has been on how cavities influence the separation sensitivity of exciton dynamics, the analysis also shows that, in the near field, thin cavities tend to reduce the coupling between dipoles as compared with free space. This implies that the rates of RET, ETU and EP will also tend to be lower. However, a special cavity setting has been identified in which sufficiently thin cavities can be used to increase one component of the dipole-dipole coupling at sufficiently large separation. For instance, a factor of ten increase in $V_{zz}$ can be obtained for intermediate zone separations with a planar cavity height of $L = 0.1\pi/p$. For a 1 eV exciton, this corresponds to a cavity height of 60 nm. Since RET and ETU rates are based on the square of the dipole coupling, they could be increased by a factor of 100. EP varies as the fourth power of the coupling and so would be enhanced by a factor of 10,000. An analogous enhancement is possible within a rectangular channel cavity by narrowing one of the cavity directions, while the second cavity dimension may be tuned to be resonant with a particular wavelength thereby enhancing emission through the Purcell effect. The result would be a cavity that  benefits both from enhanced emission and enhanced energy transfer.
 
\section{ACKNOWLEDGEMENTS}
We are pleased to be able to acknowledge useful discussions with Kaijie Xu and Ramy Ganainy who conveyed key insights into dipole-dipole interactions within optical cavities. David Andrews and David Bradshaw also provided important feedback. Interactions with Paul Martin were also helpful in the evaluating integrals associated with the planar cavity.


\vskip .3 in



{\bf\appendix{APPENDIX A: Derivation of $V_{xx}$ for Planar Cavity}}

The $V_{xx}$ component of the electric dipole coupling tensor is derived for the planar cavity of Fig. \ref{Geometry}(a) consisting of two perfectly conducting plates located at $z = 0$ and $z = L$. Without loss of generality, coordinate axes are chosen so that the donor and acceptor lie in the x,z-plane, and the unit planar wavevector, $\hat{k}_{\gamma}$, is described by its orientation in the x,y-plane: $\{{\rm cos}(\theta), {\rm sin}(\theta)\}$. Eqs. (\ref{eq:phi_Planar}) and the planar quantization expression presented in the text are applied to Eq. (\ref{eq:Vij}) to give:
\begin{multline}
V_{xx}(p,\mathbf{r}_A,\mathbf{r}_D) = \sum_{\mathbf{k}}\frac{k}{\Omega \varepsilon}\left(
\frac{{\rm sin}(\theta){\rm sin}(k_z z_A)e^{i k_\gamma x_A{\rm cos}(\theta)}{\rm sin}(\theta){\rm sin}(k_z z_D)e^{-i k_\gamma x_D{\rm cos}(\theta)}}{p-k} \right. \\
+ \frac{{\rm sin}(\theta){\rm sin}(k_z z_A)e^{-i k_\gamma x_A{\rm cos}(\theta)}{\rm sin}(\theta){\rm sin}(k_z z_D)e^{i k_\gamma x_D{\rm cos}(\theta)}}{-p-k} \\
+ \frac{\frac{i k_z}{k}{\rm cos}(\theta){\rm sin}(k_z z_A)e^{i k_\gamma x_A\frac{-i k_z}{k}{\rm cos}(\theta)}{\rm cos}(\theta){\rm sin}(k_z z_D)e^{-i k_\gamma x_D{\rm cos}(\theta)}}{p-k} \\
\left. + \frac{\frac{-i k_z}{k}{\rm cos}(\theta){\rm sin}(k_z z_A)e^{-i k_\gamma x_A{\rm cos}(\theta)}\frac{i k_z}{k}{\rm cos}(\theta){\rm sin}(k_z z_D)e^{i k_\gamma x_D{\rm cos}(\theta)}}{-p-k}\right)
\label{eq:Vxx1Planar}
\end{multline}
Define $X = x_A - x_D$, replace $\sum_{\mathbf{k}}$ with $\frac{A}{(2\pi)^2}\sum_{k_z}\int d\mathbf{k_\gamma}$, and simplify the result to give:
\begin{multline}
V_{xx}(p,X,z_A,z_D) = \frac{1}{(2\pi)^2\varepsilon L}\sum_{k_z} {\rm sin}(k_z z_A) {\rm sin}(k_z z_D)\int_0^{\infty}\sqrt{k_z^2+k_\gamma^2}\biggl( \biggr. \\
\frac{\sqrt{k_z^2+k_\gamma^2}\int_0^{2\pi}(e^{i k_\gamma X{\rm cos}(\theta)}+e^{-i k_\gamma X{\rm cos}(\theta)}){\rm sin}^2(\theta)d\theta}{p^2-(k_z^2+k_\gamma^2)} \\
+ \frac{p\int_0^{2\pi}(e^{i k_\gamma X{\rm cos}(\theta)}-e^{-i k_\gamma X{\rm cos}(\theta)}){\rm sin}^2(\theta)d\theta}{p^2-(k_z^2+k_\gamma^2)} \\
+ \frac{k_z^2}{k_z^2+k_\gamma^2}\biggl( \biggr.\frac{\sqrt{k_z^2+k_\gamma^2}\int_0^{2\pi}(e^{i k_\gamma X{\rm cos}(\theta)}+e^{-i k_\gamma X{\rm cos}(\theta)}){\rm cos}^2(\theta)d\theta}{p^2-(k_z^2+k_\gamma^2)} \\
+ \frac{p\int_0^{2\pi}(e^{i k_\gamma X{\rm cos}(\theta)}-e^{-i k_\gamma X{\rm cos}(\theta)}){\rm cos}^2(\theta)d\theta}{p^2-(k_z^2+k_\gamma^2)}\biggr. \biggl)
\biggl. \biggr) k_{\gamma} dk_{\gamma} .
\label{eq:Vxx2Planar}
\end{multline}
The angular integrals can be evaluated immediately:
\begin{align*}
\int_0^{2\pi} e^{\pm i k_\gamma X{\rm cos}(\theta)}{\rm cos}^2(\theta) d\theta &= -2\pi \frac{\partial_X^2}{k_\gamma^2}J_0(k_\gamma |X|)\\
\int_0^{2\pi} e^{\pm i k_\gamma X{\rm cos}(\theta)}{\rm sin}^2(\theta) d\theta &= 2\pi\left(1+\frac{\partial_X^2}{k_\gamma^2}\right)J_0(k_\gamma |X|) 
\end{align*}
where J$_0$ is the zeroth-order Bessel function of the first kind.  The above integral identities demonstrate that, while $X$ was not initially assumed to be positive, the expression now only depends on the magnitude of $X$. As such $X$ will now be re-defined as $X = |x_A - x_D|$. Substitution of these expressions into Eq. \eqref{eq:Vxx2Planar} and simplifying gives:
\begin{multline}
V_{xx}(p,\mathbf{r}^{(A)},\mathbf{r}^{(D)})=\frac{1}{\pi\varepsilon L}\sum_{k_z} {\rm sin}(k_z z_A) {\rm sin}(k_z z_D)\\
\left((k_z^2+\partial_X^2)\int_0^{\infty}\frac{k_\gamma J_0(k_\gamma X)}{p^2-(k_z^2+k_\gamma^2)} dk_\gamma+\int_0^{\infty}\frac{k_\gamma^3 J_0(k_\gamma X)}{p^2-(k_z^2+k_\gamma^2)} dk_\gamma\right) .
\label{eq:Vxx3Planar}
\end{multline}
The second integral can be re-cast into an analytically tractable form by applying the identity $k_\gamma^2 J_0(k_\gamma X)=-\nabla_X^2 J_0(k_\gamma X)$ and then reversing the order of integration and differentiation. The integrals can then be non-dimensionalized by introducing the following quantities: $u := k_\gamma X$, $v^2 := (p^2-k_z^2)X^2$, and $w^2 := (k_z^2-p^2)X^2$. 

To proceed further, though, the sum over $k_z$ must be broken into two regimes: $k_z < p$ and $k_z > p$:
\begin{multline}
V_{xx}(p,\mathbf{r}^{(A)},\mathbf{r}^{(D)})=\frac{-1}{\pi\varepsilon L}\left(\sum_{k_z<p} {\rm sin}(k_z z_A) {\rm sin}(k_z z_D)
\left( (k_z^2+\partial_X^2-\nabla_X^2)\int_0^{\infty}\frac{u J_0(u)}{u^2-v^2} du\right) +\right. \\
\left. \sum_{k_z>p} {\rm sin}(k_z z_A) {\rm sin}(k_z z_D)
\left( (k_z^2+\partial_X^2-\nabla_X^2)\int_0^{\infty}\frac{u J_0(u)}{u^2+w^2} du\right) \right) .
\label{eq:Vxx4Planar}
\end{multline}
The first integral diverges at $u = v$, so this is resolved by introducing an $i\epsilon$-prescription. It also turns out to be useful to re-express the Bessel functions in terms of Hankel functions:
\begin{multline}
V_{xx}(p,\mathbf{r}^{(A)},\mathbf{r}^{(D)})=\frac{-1}{2\pi\varepsilon L}\biggl( \biggr.\sum_{k_z<p} {\rm sin}(k_z z_A) {\rm sin}(k_z z_D)\\
\left( (k_z^2+\partial_X^2-\nabla^2)\left( \lim_{\epsilon\rightarrow 0} \int_0^{\infty}\frac{u H_0^{(1)}(u)}{u^2-(v\pm i\epsilon)^2} du+\int_0^{\infty}\frac{u H_0^{(2)}(u)}{u^2-(v\pm i\epsilon)^2} du\right)\right) +\\
\sum_{k_z>p} {\rm sin}(k_z z_A) {\rm sin}(k_z z_D)
\left( (k_z^2+\partial_X^2-\nabla^2)\left( \int_0^{\infty}\frac{u H_0^{(1)}(u)}{u^2+w^2} du+\int_0^{\infty}\frac{u H_0^{(2)}(u)}{u^2+w^2} du\right)\right)\biggl.\ \biggr)
\label{eq:Vxx5Planar}
\end{multline}
Each pair of integrals can be evaluated using complex contour integration about a path along the first quadrant of the complex u-plane for the Hankel functions of the first kind, and along the fourth quadrant for the Hankel functions of the second kind. The selection of $\pm i\epsilon$ leads to two separate answers that differ only in phase. Within the perturbation theory setting, this difference in phase is unimportant, and so the $+i\epsilon$ term will be selected~\cite{Jenkins2003}:
\begin{multline}
V_{xx}(p,\mathbf{r}^{(A)},\mathbf{r}^{(D)})=\\
\frac{-1}{2\pi\varepsilon L}\biggl( \biggr.\sum_{k_z<p} {\rm sin}(k_z z_A) {\rm sin}(k_z z_D)
\left( (k_z^2+\partial_X^2-\nabla_X^2)\left(\pi i H_0^{(1)}(v) \right)\right) +\\
\sum_{k_z>p} {\rm sin}(k_z z_A) {\rm sin}(k_z z_D)
\left( (k_z^2+\partial_X^2-\nabla_X^2)\left( 2 K_0(w) \right)\right) \biggl. \biggr) .
\label{eq:Vxx6Planar}
\end{multline}

Writing out the expressions for $u$ and $w$, we have that
\begin{multline}
V_{xx}(p,\mathbf{r}^{(A)},\mathbf{r}^{(D)})=\\
\frac{-1}{2\pi\varepsilon L}\sum_{k_z<p} {\rm sin}(k_z z_A) {\rm sin}(k_z z_D)
 (k_z^2+\partial_X^2-\nabla_X^2)\bigl( \pi i H_0^{(1)}(X\sqrt{p^2-k_z^2}) \bigr) \\
- \frac{1}{2\pi\varepsilon L} \sum_{k_z>p} {\rm sin}(k_z z_A) {\rm sin}(k_z z_D)
(k_z^2+\partial_X^2-\nabla_X^2)\left( 2 K_0(X\sqrt{k_z^2-p^2}) \right) .
\label{eq:Vxx7Planar}
\end{multline}
This result can be written more compactly by noting that $K_0(w) = \frac{\pi i}{2}H_0^{(1)}(i w)$ so that
\begin{equation}
V_{xx}(p,\mathbf{r}^{(A)},\mathbf{r}^{(D)})=\frac{-i}{2\varepsilon L}\sum_{k_z} {\rm sin}(k_z z_A) {\rm sin}(k_z z_D)
\left( (k_z^2+\partial_X^2-\nabla_X^2)H_0^{(1)}(X\sqrt{p^2-k_z^2}) \right) .
\label{eq:Vxx8Planar}
\end{equation}
Finally the differential operators are evaluated to give
\begin{multline}
V_{xx}(p,X,z_A,z_D)=\frac{-i}{4\varepsilon L}\sum_{k_z} {\rm sin}(k_z z_A) {\rm sin}(k_z z_D)\\
\left( \left(p^2+k_z^2\right) H_0^{(1)}\left( X\sqrt{p^2-k_z^2}\right)+\left( p^2-k_z^2\right)H_2^{(1)}\left(X\sqrt{p^2-k_z^2}\right) \right) .
\label{eq:Vxx8Planar}
\end{multline}

Analogous expressions for the other components of the dipole coupling tensor are derived in the Supporting Information.

\vskip .3 in

{\bf \appendix{APPENDIX B: Derivation of $V_{xx}$ for Channel Cavity}}

The $V_{xx}$ component of the electric dipole coupling tensor is derived for the channel cavity of Fig. \ref{Geometry}(b) consisting of four perfectly conducting plates with a rectangular cross-section. We start by substituting the electric modes of Eq. (\ref{eq:phi_Channel}) and the quantization condition of Eq. (\ref{eq:E0_Channel}) into Eq. (\ref{eq:Vij}) to give:
\begin{multline}
V_{xx}(p,\mathbf{r}_A,\mathbf{r}_D)=\sum_{\mathbf{k}}\frac{2}{\Omega \varepsilon}\frac{k_\eta^2}{k}\\
\left(\frac{{\rm sin}(k_y y_A){\rm sin}(k_z z_A)e^{i k_x x_A}{\rm sin}(k_y y_D){\rm sin}(k_z z_D)e^{-i k_x x_D}}{p-k} \right. \\
\left. + \frac{{\rm sin}(k_y y_A){\rm sin}(k_z z_A)e^{-i k_x x_A}{\rm sin}(k_y y_D){\rm sin}(k_z z_D)e^{i k_x x_D}}{-p-k}\right) .
\label{eq:Vxx1WG}
\end{multline}
Define $X = x_A - x_D$, replace $\sum_{\mathbf{k}}$ with $\frac{L}{(2\pi)}\sum_{k_y,k_z}\int d k_x$, and simplify the result to give:

\begin{multline}
V_{xx}(p,\mathbf{r}_A,\mathbf{r}_D)=\frac{2}{\pi\varepsilon a b}\sum_{k_y,k_z}
{\rm sin}(k_y y_A) {\rm sin}(k_z z_A){\rm sin}(k_y y_D) {\rm sin}(k_z z_D)\\
\int_{-\infty}^{\infty}\frac{k_\eta^2}{k_x^2+k_\eta^2}\frac{\sqrt{k_x^2+k_\eta^2} {\rm cos}(k_x X)+i p{\rm sin}(k_x X)}{p^2-k_x^2-k_\eta^2}d k_x .
\label{eq:Vxx2WG}
\end{multline}
It is useful to introduce quantities with which the integral above can be non-dimensionalized: $u := k_x X$, $v^2 := (p^2-k_\eta^2) X^2$, $w^2 := (k_\eta^2-p^2) X^2$, and $t := k_\eta X$. In terms of these quantities,
\begin{multline}
V_{xx}(p,\mathbf{r}_A,\mathbf{r}_D)=\frac{2}{\pi\varepsilon a b X}\left(\left(\sum_{k_\eta<p}
{\rm sin}(k_y y_A) {\rm sin}(k_z z_A){\rm sin}(k_y y_D) {\rm sin}(k_z z_D)\right.\right.\\
\left.\frac{i t^2}{p X}\int_{-\infty}^{\infty}\frac{{\rm sin}(u)}{\sqrt{u^2+t^2}}d u
-\frac{i t^2}{p X}\int_{-\infty}^{\infty}\frac{\sqrt{u^2+t^2}}{u^2-v^2}{\rm sin}(u)d u
-t^2\int_{-\infty}^{\infty}\frac{{\rm cos}(u)}{u^2-v^2}d u\right) \\
+ \left(\sum_{k_\eta>p}
{\rm sin}(k_y y_A) {\rm sin}(k_z z_A){\rm sin}(k_y y_D) {\rm sin}(k_z z_D)\right.\\
\left.\left.\frac{i t^2}{p X}\int_{-\infty}^{\infty}\frac{{\rm sin}(u)}{\sqrt{u^2+t^2}}d u
-\frac{i t^2}{p X}\int_{-\infty}^{\infty}\frac{\sqrt{u^2+t^2}}{u^2+w^2}{\rm sin}(u)d u
-t^2\int_{-\infty}^{\infty}\frac{{\rm cos}(u)}{u^2+w^2}d u\right)\right) .
\label{eq:Vxx3WG}
\end{multline}
The sine integrals are odd and therefore integrate to zero. The divergent nature of the first integral diverges is resolved using an $i\epsilon$ prescription. In addition, it is profitable to express both cosine integrals in terms of complex exponentials. Then
\begin{multline}
V_{xx}(p,\mathbf{r}_A,\mathbf{r}_D)=\frac{-1}{\pi\varepsilon a b X}
\left(\left(\sum_{k_\eta<p}{\rm sin}(k_y y_A) {\rm sin}(k_z z_A){\rm sin}(k_y y_D) {\rm sin}(k_z z_D)\right.\right.\\
\left.t^2\left( \lim_{\epsilon\rightarrow 0}\int_{-\infty}^{\infty}\frac{e^{i u}}{u^2-(v\pm i\epsilon)^2}d u+\int_{-\infty}^{\infty}\frac{e^{-i u}}{u^2-(v\pm i\epsilon)^2}d u\right)\right)+\\
\left.\left(\sum_{k_\eta>p}{\rm sin}(k_y y_A) {\rm sin}(k_z z_A){\rm sin}(k_y y_D) {\rm sin}(k_z z_D)
t^2\left(\int_{-\infty}^{\infty}\frac{e^{i u}}{u^2+w^2}d u+\int_{-\infty}^{\infty}\frac{e^{-i u}}{u^2+w^2}d u\right)\right)\right).
\label{eq:Vxx4WG}
\end{multline}
Each pair of integrals can be evaluated using complex contour integration about a path along the upper half plane for the positive exponential integrals, and in the lower half plane for the negative exponential integrals. The selection of $\pm i\epsilon$ leads to two separate answers that differ only by a difference in phase. When using perturbation theory this difference in phase is unimportant,  so the $+i\epsilon$ term will be selected~\cite{Jenkins2003}. As with the planar cavity, $X$ was not initially assumed to be either to be positive, but it becomes positive by virtue of an integral. Once again, $X$ will now be re-defined as $X = |x_A - x_D|$. The final expression is:
\begin{equation}
V_{xx}(p,\mathbf{r}_A,\mathbf{r}_D)=\frac{-2 i}{\varepsilon a b}
\sum_{k_y,k_z}{\rm sin}(k_y y_A) {\rm sin}(k_z z_A){\rm sin}(k_y y_D){\rm sin}(k_z z_D)
\left(\frac{k_\eta^2}{\sqrt{p^2 - k_\eta^2}}e^{i X\sqrt{p^2 - k_\eta^2}}\right) .
\label{eq:Vxx5WG_app}
\end{equation}

\end{document}